\documentclass[%
 aip,
 amsmath,amssymb,
 reprint,%
]{revtex4-1}

\usepackage{amsmath}
\usepackage[dvips]{graphicx}
\usepackage{epsfig}
\usepackage{tabularx}
\usepackage{color}
\usepackage[pdfpagemode=UseNone,colorlinks=true,linkcolor=blue,citecolor=blue,urlcolor=blue]{hyperref}

\usepackage{soul}
\usepackage{gensymb}

\begin{document}
\title[]{Super-resolution imaging of a low frequency levitated oscillator}

\author{N. P. Bullier}%
 \email{nathanael.bullier.15@ucl.ac.uk}
\affiliation{Department of Physics and Astronomy, University College London, Gower Street, London WC1E 6BT, United Kingdom}%

\author{A. Pontin}
 \email{a.pontin@ucl.ac.uk}
\affiliation{Department of Physics and Astronomy, University College London, Gower Street, London WC1E 6BT, United Kingdom}

\author{P. F. Barker}
 \email{p.barker@ucl.ac.uk}
\affiliation{Department of Physics and Astronomy, University College London, Gower Street, London WC1E 6BT, United Kingdom}%

\begin{abstract}
We describe the measurement of the secular motion of a levitated nanoparticle in a Paul trap with a CMOS camera. This simple method enables us to reach signal-to-noise ratios as good as 10$^{6}$ with a displacement sensitivity better than 10$^{-16}$\,m$^2$/Hz. This method can be used 
to extract trap parameters as well as the properties of the levitated particles. We demonstrate continuous monitoring of the particle dynamics on timescales of the order of weeks. We show that by using the improvement given by super-resolution imaging, a significant reduction in the noise floor can be attained, with an increase in the bandwidth of the force sensitivity. This approach represents a competitive alternative to standard optical detection for a range of low frequency oscillators where low optical powers are required.

\end{abstract}

\maketitle

\section{Introduction}
Levitated nanoparticle oscillators are seen as promising platforms for exploring the macroscopic limits of quantum mechanics. For example, levitated nanoparticles containing a single spin are potential candidates for creating a macroscopic superposition via a type of Ramsey interferometry \cite{Sougato1}. Motional cooling of nanoparticles within a trap, followed by a controlled release through a matter wave diffraction grating, also offers the potential for creating and evidencing macroscopic superposition \cite{Bateman2014}. More recently, non-interferometric methods have been proposed as a way of testing wavefunction collapse models by measuring the excess noise in the trapped motion of a levitated nanoparticle \cite{Goldwater}. 


A distinguishing feature of many of these levitated oscillators is low oscillation frequency, typically in the range of a few Hz to kHz. Detection of small amplitude mechanical motion at these low frequencies is challenging as environmental mechanical noise in the detection chain can swamp measurements of the trapped motion of the nanoparticle.  A common approach to measurement of particle displacement uses a difference detection scheme where a laser beam is used to illuminate the particle. The transmitted light modified by scattering from the particle is typically directed onto a quadrant photodiode or alternatively split into two equal components using a mirror such that each component is detected on a separate photodiode. The motion of the particle is observed as an imbalance in the difference between the currents on the photodiodes \cite{Gieseler}. While this is relatively noise free at frequencies exceeding 10\,kHz, low frequency mechanical noise can induce beam pointing noise which cannot easily be differentiated from the true mechanical motion of the oscillator.       
\par We present a very simple and surprisingly sensitive imaging method for measuring low frequency particle displacement which is free of the low mechanical noise observed in split detection methods but yet demonstrates displacement sensitivity of better than 10$^{-16}$\,m$^2$/Hz.  

\begin{figure}[!ht]
\includegraphics[width=8.6cm]{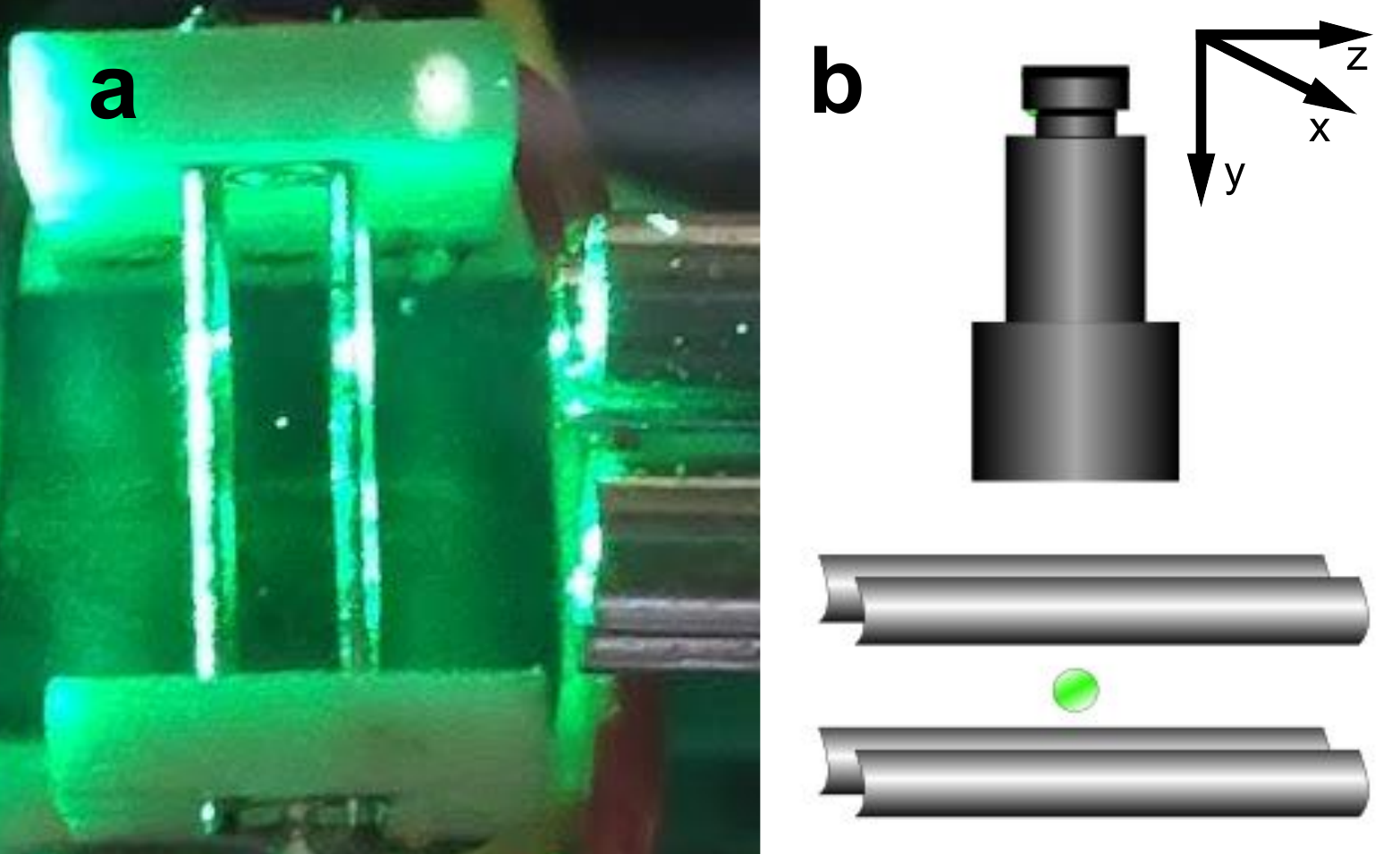}
\caption{(a) An image of the trap with a nanoparticle in its centre. Image taken from the camera axis (\textit{y}-axis). The end of the quadrupole guide can be seen on the right-hand side of the picture. It is used to load the trap and as a quadrupole mass filter to select a given charge-to-mass ratio. (b) A schematic view with camera position and axes defined. The laser propagates along the \textit{x}-axis. The trap parameters discussed in the main text are the following: $r_o=1.1$\,mm, $z_o=3.5$\,mm, $\eta=0.815/2$ (asymmetric driving), $\kappa=0.086$, and $\epsilon=0.5$.}
\label{fig1}
\end{figure}

\section{Levitation in a Paul trap}
We use a single nanoparticle trapped in a linear Paul trap to create a levitated nanoparticle oscillator whose trap frequency is dependent on the surface charge and the applied voltage on the electrodes of the trap. A picture of the trap is shown in Fig.\,\ref{fig1}(a) which consists of four electrodes placed at the corner of a square to provide an AC field that confines the nanoparticle in the \textit{x-y} plane with two additional electrodes to apply a DC voltage that traps the motion along the \textit{z}-axis. In this configuration, the total electric field near the center of the trap can be written as \cite{wineland}:


\begin{equation}\label{eq1}
\begin{split}
\hat{E}(x,y,z,t)=&-2 \eta V_o \left(\frac{x \hat{x}-y \hat{y}}{r_o^{2}} \right) \text{cos}(\omega_{d} t)\\&-\frac{2\kappa U_o}{z_o^2}\left(z \hat{z}-\epsilon x \hat{x}-(1-\epsilon) y \hat{y} \right)
\end{split}
\end{equation}

\noindent where $r_o$ and $z_o$ are the distances between the trap center and the AC and DC electrodes respectively, $V_o$ and $U_o$ are the applied potentials, $\omega_{d}$ is the AC drive frequency, $\eta$ and $\kappa$ are geometrical efficiency coefficients\cite{efficiency1,efficiency2} quantifying non-perfectly quadratic potentials and $\epsilon$ is the trap ellipticity. We obtain the efficiency parameters with numerical FEM simulations (see caption of Fig.\,\ref{fig1} for the numerical values). We drive the trap asymmetrically by applying the same AC potential on two opposite electrodes and grounding the other two.


The AC potential gives rise to a pseudo-potential $\frac{q }{4m\omega_{d}^{2}}|\nabla V(x,y)|^2$ with $V(x,y)=V_o\left(\frac{1}{2}+\eta\frac{x^{2}-y^{2}}{r_o^{2}} \right)$ and $q/m$ the charge-to-mass ratio. Following from the expression of the electric field given in Eq.\,\ref{eq1}, one can define stability parameters \cite{wineland} $a_{i}$ and $q_{i}$ corresponding to the motion along the $i$-axis given by the AC and DC potential respectively:

\begin{equation}\label{eq2}
\begin{split}
a_{x}=a_{y}=-\frac{1}{2}a_{z}=-\frac{q}{m}\frac{4\kappa\,U_{o}}{z_{o}^{2}\,\omega_{d}^{2}} \\
q_{x}=-q_{y}=\frac{q}{m}\frac{4\eta V_{o}}{r_{o}^{2}\,\omega_{d}^{2}},\,q_{z}=0
\end{split}
\end{equation}

\noindent where we assumed $\epsilon=0.5$. In the case where $a_{i}\ll1$, $q_{i}\ll1$, the pseudo-potential gives rise to secular frequencies which can be approximated to $\omega_{i}\approx\frac{\omega_{d}}{2}\sqrt{a_{i}+\frac{1}{2}q_{i}^{2}}$.


The trap is loaded with nanoparticles in low vacuum, at a pressure of $\sim 10^{-1}$\,mbar, by means of electrospray ionization\cite{electrospray}, where the nanoparticles are suspended in ethanol with a concentration $\sim$\,10\,$\mu$g/mL. In order to increase the flux of particles reaching the trapping region, a quadrupole guide ``connects" the output skimmer and the Paul trap which are placed roughly $20$\,cm apart. Its end section can be seen in Fig.\,\ref{fig1}(a). The guide is usually operated in a mass filter configuration \cite{march}, where a high DC voltage is applied onto two electrodes such that only the tip of the first stability region of the guide is allowed. Note that since the trap and the guide have different geometries, the effective stability region is further reduced. We typically trap particles with charge-to-mass ratios in the range  $0.05<q/m<2$\,C/kg.


\section{Measuring displacement via imaging}

Optical interferometric detection schemes often offer the best sensitivities in terms of displacement. However, typical sensitivities are often much worst at low frequency. For example, the noise floor of a standard balanced Michelson interferometer can be of the order of $10^{-30}$\,m$^2$/Hz around $100$\,kHz but it can easily drop to $10^{-20}$\,m$^2$/Hz at $100$\,Hz due to environmental noise affecting the beam paths as well as flicker noise \cite{Slezak2018}. This difference can be even more pronounced when dealing with nanoparticle motion due to the intrinsically lower coupling/detection efficiency.

\begin{figure}[!ht]
\includegraphics[width=8.6cm]{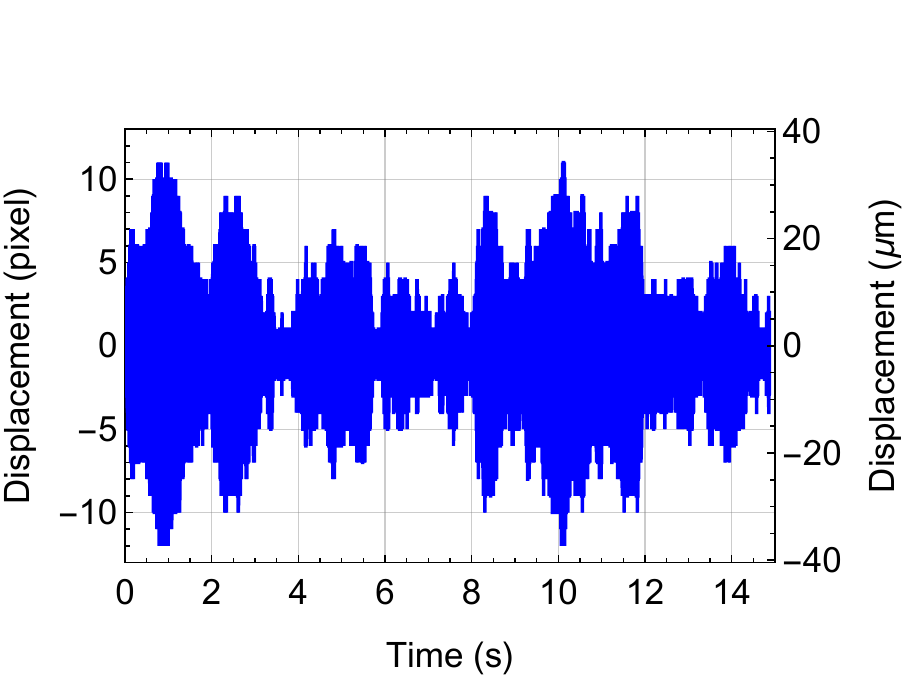}
\caption{Time-traces of the motion along the \textit{z}-axis of a silica nanoparticle of mass $9.6\times10^{-17}$\,kg at 10$^{-3}$\,mbar, derived from imaging. The particle position is obtained by considering the pixel of highest intensity. The displacement is calibrated by moving the camera on a translational stage. The resulting calibration coefficient is (3.11$\,\pm\,0.03)\,\mu$m/pixel.}
\label{fig2}
\end{figure}

Here, we exploit a much simpler detection scheme. The trapped nanoparticle is illuminated with a green laser diode with powers ranging from 10\,mW to 40\,mW and a beam waist of 250\,$\mu$m. The light scattered at $90$\degree\, is collected by a zoom objective \cite{objective} mounted on a lens \cite{lens} and imaged by a low cost CMOS camera \cite{camera}. We extract the particle position on the \textit{x-z} plane by finding the coordinates of the pixel of highest intensity. The motion of the particle is calibrated by moving the camera by a known amount using a translational stage. An example of a time trace for a silica nanoparticle with a mass of $9.6\times10^{-17}$\,kg can be seen in Fig.\,\ref{fig2}. The camera sensor has a resolution of 1280\,x\,1024, however a smaller area of interest can be addressed, allowing an acquisition at a faster frame rate. In our typical configuration we exploit a 50\,x\,40 matrix that allows to acquire at $\sim800$\,fps. This, in combination with the simplicity of the algorithm to reconstruct the particle position, allows for a real time acquisition of two time traces corresponding to the particle motion projected onto the coordinates defined by the camera pixel matrix. It is worth noting that the time series are composed of $8$\,bit integers. Combined with the low sampling rate allows for continuous monitoring of the particle position over very long times without having large data storage requirements.

The motion of the nanoparticle in the trap can be decomposed in two different motions. A secular motion, driven by stochastic forces, such as collisions by gas particles, and a motion at the drive frequency $\omega_{d}$. Here, typically, $\omega_{d}$ is a few kHz. The secular motion can be easily kept at frequencies smaller than $500$\,Hz. The motion can be described by a damped harmonic oscillator. The single-sided power spectral density (PSD) of the motion along any axis $\textit{i}$ can be written as: $S_{i}(\omega)=|\chi(\omega)|^{2} (S_{F_{th}}+S_{F})$ where $S_{F_{th}}=4k_{B}Tm\Gamma$ is the PSD of the thermal Langevin force while $S_{F}$ is the PSD of the total force noise due to all other unknown sources. The bath temperature is given by $T$, $k_{B}$ is the Boltzmann constant, $m$ the mass of the nanoparticle and $\Gamma$ the viscous damping due to collisions with residual gas. Finally, $\chi(\omega)$, is the mechanical susceptibility, given by $\chi(\omega)= (m(\omega_{i}^{2}-\omega^{2}-i\Gamma\omega))^{-1}$ where $\omega_{i}$ is the resonant frequency which here corresponds to the secular frequency along the \textit{i}$^{th}$ axis.

\begin{figure}[!ht]
\includegraphics[width=8.6cm]{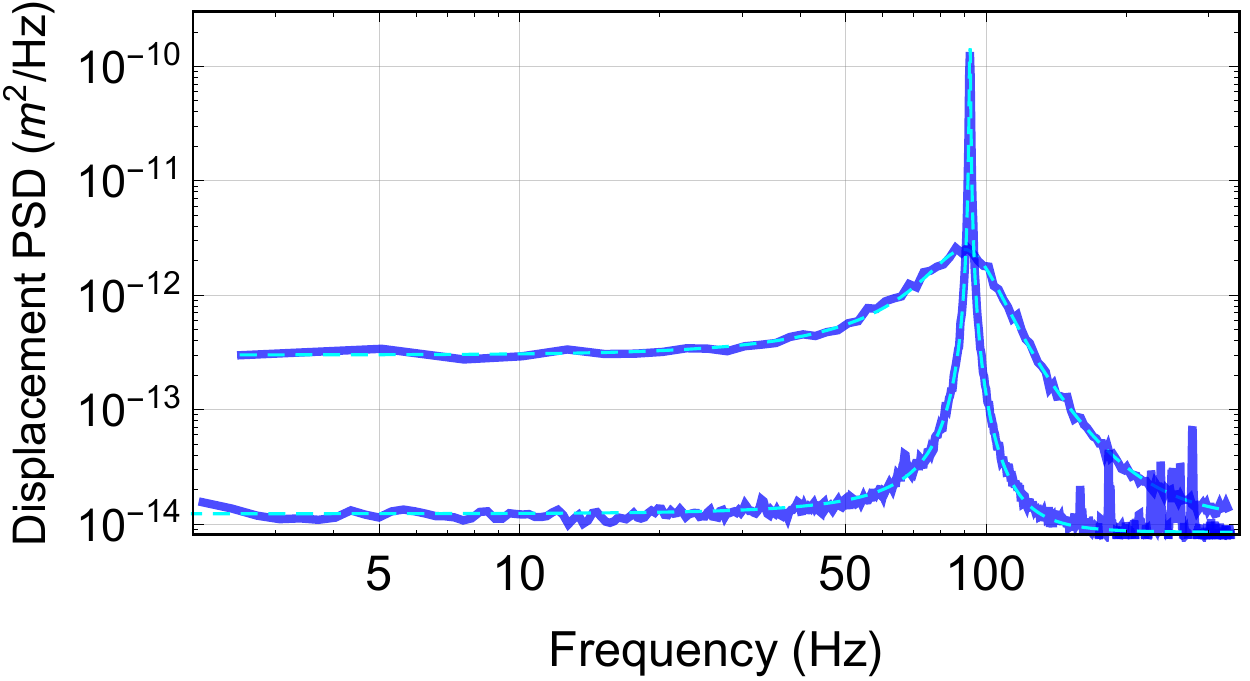}
\caption{Calibrated Power Spectral Densities (PSD) of the motion of a trapped nanoparticle at 0.10\,mbar and 1.0$\times$10$^{-3}$\,mbar. Blue continuous lines are experimental PSD of the motion along the \textit{z}-axis. The secular frequency is 92.5\,Hz. The broader linewidth corresponds to the motion at 0.10\,mbar. The fits of the mechanical susceptibilities are shown in dashed cyan. The time trace shown in Fig.\,\ref{fig2} is a subsection of the timetrace used to obtain the spectra shown at 1.0$\times$10$^{-3}$\,mbar.}
\label{fig3}
\end{figure}

We show in Fig.\,\ref{fig3} typical spectra for the same nanoparticle as the one shown in Fig.\,\ref{fig2} taken at different pressures along the \textit{z}-axis. Despite the very simple approach, this method provides spectra with high signal-to-noise ratios (SNR). Along with the experimental spectra, we show fit results assuming a total force noise PSD. Down to pressures of the order of $\sim 10^{-3}$\,mbar the spectral noise floor are sufficiently low to resolve the susceptibility down to the zero frequency limit.

\begin{figure}[!ht]
\includegraphics[width=8.6cm]{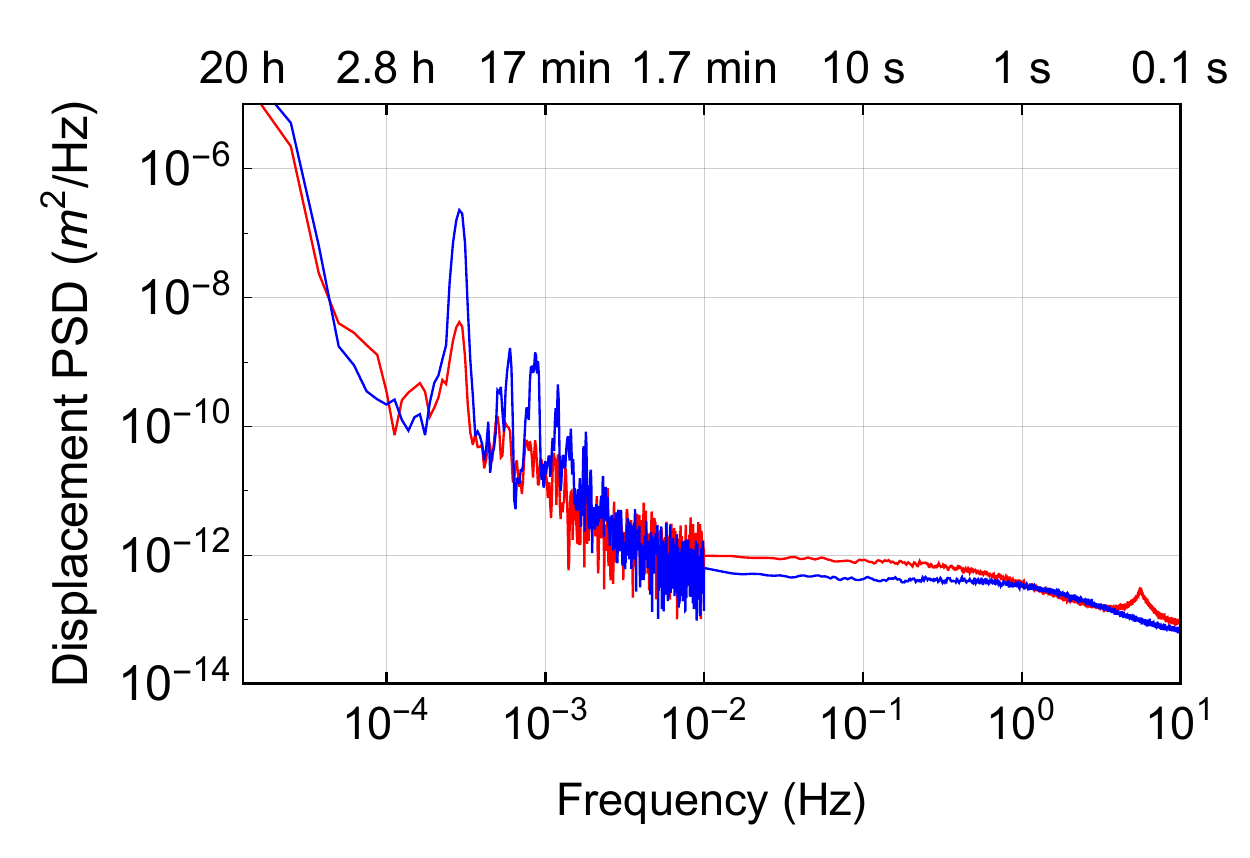}
\caption{Low frequency PSDs calculated from data acquired for almost three days. Red and blue lines represent the PSD along the \textit{x} and \textit{z}-axis respectively. The peak around $1$\,hour corresponds to motion in the trap correlated to temperature fluctuations in the lab. At frequencies above 10$^{-2}$\,Hz, the PSD has been averaged with more spectra, which leads to the smoother PSD profiles in this range.}
\label{fig4}
\end{figure}

The continuous monitoring  of the particle motion, allows us to explore the dynamics down to extremely low frequencies. Taking into consideration the longest continuous stretch of data that lasts almost $3$\,days we obtain the displacement PSD shown in Fig.\,\ref{fig4}. Interestingly, a peak (and its harmonics) is clearly visible corresponding to a period of $\sim1$\,hour. It is possible to show that this displacement modulation is highly correlated to a modulation of the secular frequency, itself correlated to the temperature drifts in the lab. Indeed, the temperature induced modulation of the drive signal amplitude is $\sim1$\,V (peak-to-peak) on the trap electrodes which completely explains the changes in the secular frequencies.


\bigskip
\section{Super-resolution measurements}
The spectra shown so far in Figures \ref{fig3} and \ref{fig4} have been calculated on time traces where the position of the particle has been determined by finding the pixel of highest intensity. This naive approach enables real time traces by controlling the camera with a Python library \cite{instrumental}. This is enough to extract basic information such as frequencies and linewidths characterizing the mechanical motion. Applications such as sensing or cooling of the centre-of-mass motion may require higher sensitivities which can be achieved by using super-resolution imaging.

\begin{figure}[!ht]
\includegraphics[width=8.6cm]{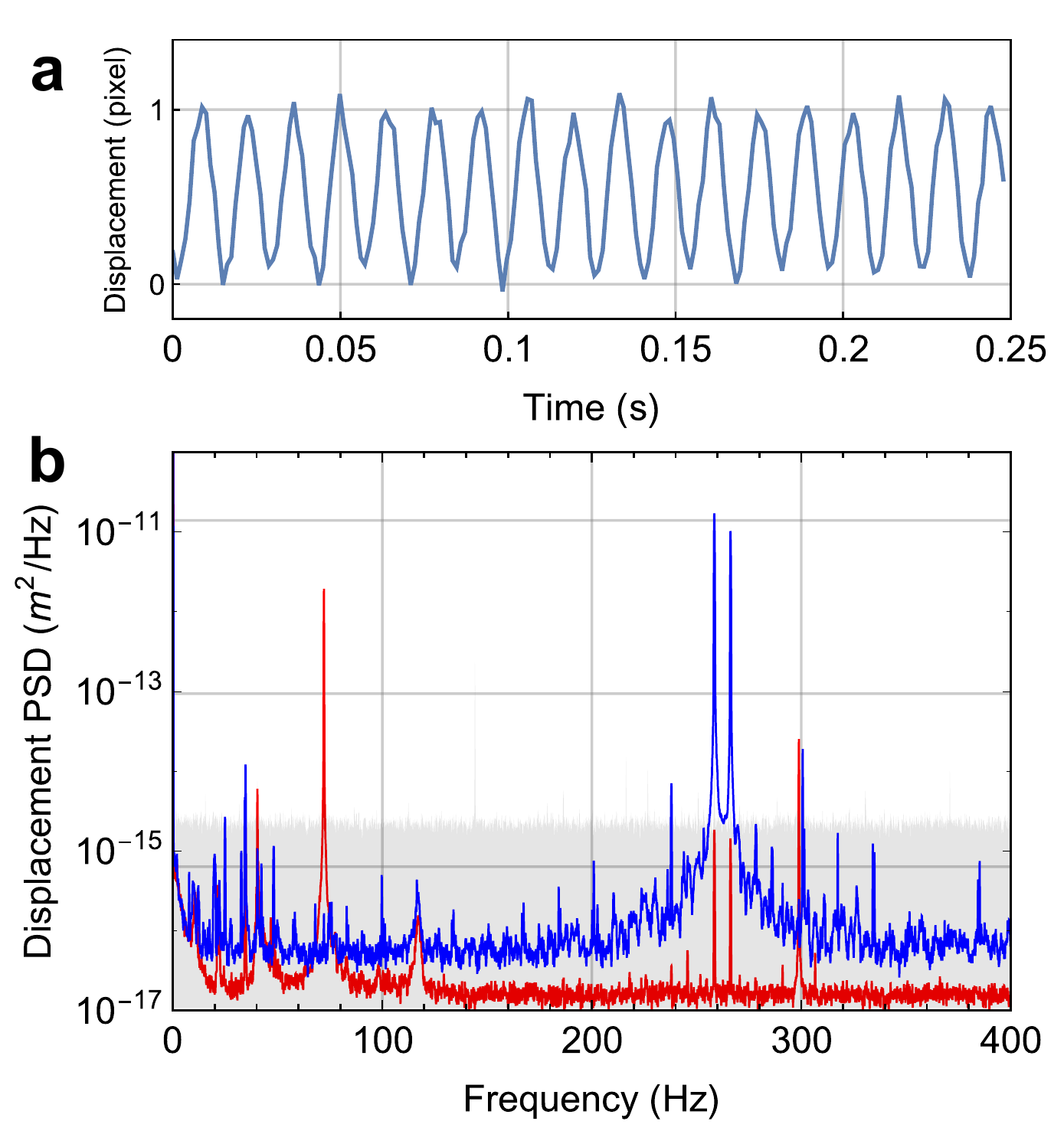}
\caption{Resolution enhancement obtained by fitting a Gaussian distribution to find the particle position on each frame. (a) Time-trace of the motion along the \textit{z}-axis obtained with the resolution enhancement showing sub-pixel resolution. (b) PSD of the particle displacement. Red and blue lines refer to the motion along the \textit{z} and \textit{x}-axis respectively. The gray shaded area shows the noise floor when the algorithm finds the intensity maximum to obtain the particle position.
}
\label{fig5}
\end{figure}

Complex algorithms to obtain sub-pixel resolution is applied in many fields including particle physics\cite{subpx1}, chemistry\cite{subpx2} and astrophysics. In the latter case, for example, it was applied to enhance the search for exoplanets\cite{subpx3}. Here, we extract the particle position by fitting a two-dimensional Gaussian profile to the spatial intensity distribution given by the camera. This increases the sensitivity by more than two orders of magnitude as it can be seen in Fig.\,\ref{fig5}, where we plot a comparison between PSDs with and without the resolution enhancement for both trap directions. These spectra were taken at a pressure $\sim 10^{-6}$\,mbar for a particle with a charge-to-mass ratio of $0.1$\,C/kg. The secular frequencies along the \textit{x}, \textit{y}, and \textit{z}-axis are $72.1$, $258.5$ and $266.2$\,Hz respectively. A trap ellipticity slightly different from the nominal one of 0.5 removes the degeneracy on the \textit{x} and \textit{y}-axes, which are easily resolved at this pressure. The peak in the PSD at $298$\,Hz corresponds to the aliasing of the excess micromotion at the AC drive frequency of $1.1$\,kHz.



\begin{figure}[!ht]
\includegraphics[width=8.6cm]{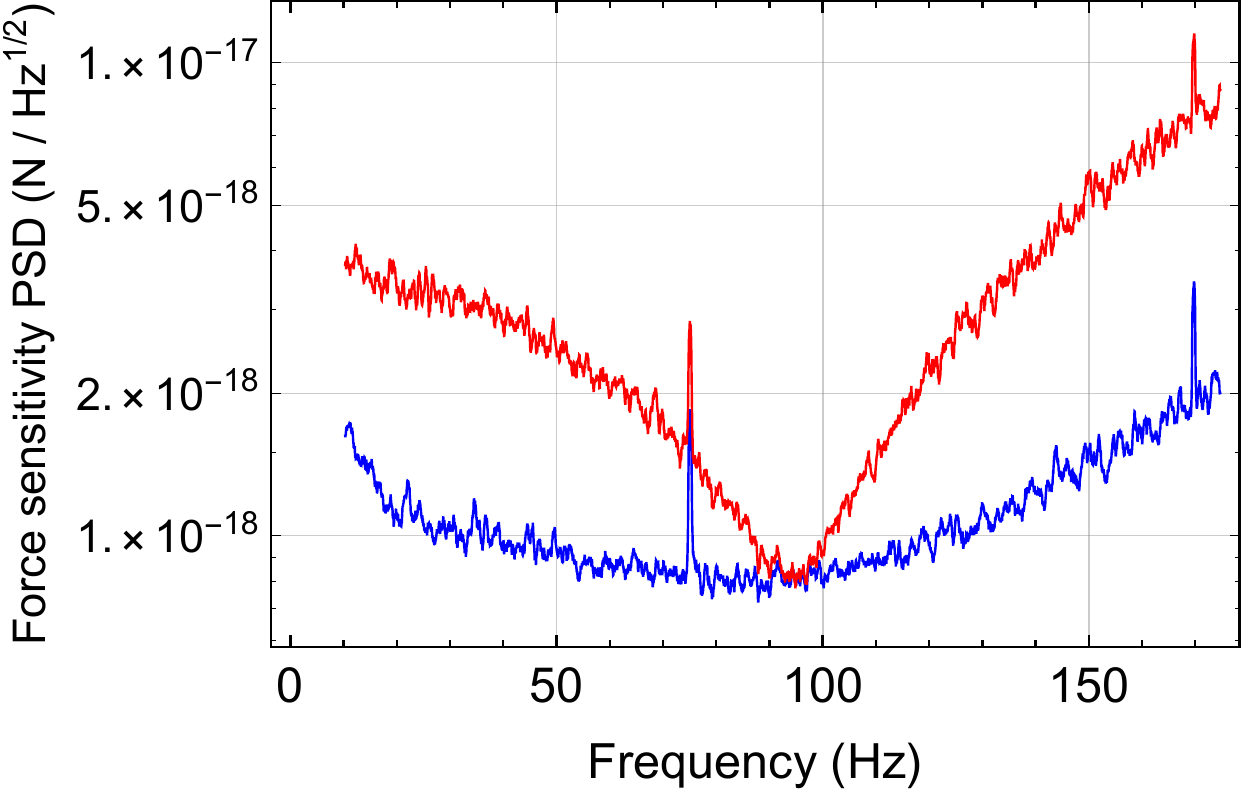}
\caption{Force-noise sensitivity of a silica nanoparticle with a mass of 9.6$\times10^{-17}$\,kg at 7.4$\times$10$^{-5}$\,mbar along the \textit{z}-axis. The red curve corresponds to the force-noise sensitivity obtained without the super-resolution enhancement. The blue one is obtained on the same acquisition but using the super-resolution. This increases the 3\,dB bandwidth, centred around 90\,Hz, from 21\,Hz to 112\,Hz.
}
\label{fig6}
\end{figure}

Given its low mass and small damping rate, a levitated nanosphere in vacuum is a sensitive force sensor. The super-resolution enhancement can be used to increase the force sensitivity bandwidth too. We show an example in Fig.\,\ref{fig6} that compares the force sensitivity of a silica nanoparticle with a mass of 9.6$\times10^{-17}$\,kg levitated at 7.4$\times$10$^{-5}$\,mbar using super-resolution and the method that finds the brightest pixel to extract the particle position. It enables us to increase the 3\,dB force-noise sensitivity bandwidth from 21\,Hz to 112\,Hz. This last one peaks at (8.2$\,\pm\,0.6)\times10^{2}$\,zN$/\sqrt{\textrm{Hz}}$. At this pressure, the force noise contribution from the Brownian motion, and therefore the lowest achievable limit without other noise, is $\sqrt{S_{F_{th}}}$=(6.3$\,\pm\,$0.3)$\times10^{2}$\,zN$/\sqrt{\textrm{Hz}}$. The force noise sensitivity, $\sqrt{S_{F}}$, shown in Fig.\,\ref{fig6} is calculated from the displacement PSD $S_{z}(\omega)$ as $\sqrt{S_{F}(\omega)}=\frac{\sqrt{S_{z}(\omega)}}{|\chi(\omega)|}$ with $\chi(\omega)$ the mechanical susceptibility defined above. The mechanical susceptibility depends on the mass, independently measured, as well as the damping and the secular frequency, obtained by fitting a displacement PSD from an independent data set. This approach provides a good estimate of the force sensitivity, however, a more rigorous analysis can be obtained exploiting Wiener filter theory \cite{wienerfiltering}.

\bigskip
\section{Conclusions}
We have demonstrated the use of a low-cost CMOS camera to make precision displacement measurements of a low frequency high-Q mechanical oscillator. Although applied to a Paul trap, this method could be used for many applications from sensing to cooling in a range of oscillators.
\par The three factors which can be improved in the measurements presented here are the sensitivity, the acquisition rate and the fitting speed of the algorithm used to determine the particle position. The sensitivity of the measurements can easily be improved by increasing the magnification. The secular frequencies were limited here to 500\,Hz by the acquisition rate. In order to directly detect higher secular frequencies and avoid aliasing, it is possible to use cameras with higher pixel clocks (i.e., $\sim500$\,MHz against $\sim40$\,MHz here), opening the possibility of getting frame rates as high as $4$\,kHz. Furthermore, fluorescence correlation spectroscopy techniques \cite{Kim2007} could be used in the case of fluorescing nanoparticles such as YLF crystals doped in Yb$^{3+}$ \cite{refrigiration}. This could lead to frame rates of tens of kHz. Recent Single-Photon Avalanche Diode (SPAD) cameras have been operated up to $300$\,kfps \cite{spad}. 

While here the time traces without subpixel resolution were live, the data with super-resolution were obtained by post-processing the video. Indeed, the fitting routine used was too slow to keep acquiring data at $\sim800$\,frames per second (fps). It is nonetheless still possible to acquire live position time traces with very high frame rates by running the fitting algorithm on a Graphics Processing Unit (GPU). Gaussian fits or better algorithms can then achieve fit speeds $\sim 100000$\,fits/s for a 10\,x\,10 pixel matrix \cite{fastfit1,fastfit2}.

\section{acknowledgments}
The authors acknowledge funding from the EPSRC Grant No. EP/N031105/1 and from the EU H2020 FET project TEQ (Grant No. 766900). NPB acknowledges funding from the EPSRC Grant No. EP/L015242/1. AP has received funding from the European Union’s Horizon 2020 research and innovation programme under the Marie Sklodowska-Curie Grant Agreement No. 749709.

\bibliography{paper_camera_v2_bib}

\begin{thebibliography}{23}
\bibitem [{len()}]{lens}%
  \BibitemOpen
  \href@noop {} {\ }\bibinfo {note} {The lens used is a fixed focal length Lens
  extender from Edmund optics}\BibitemShut {NoStop}%
\bibitem [{cam()}]{camera}%
  \BibitemOpen
  \href@noop {} {\ }\bibinfo {note} {The camera used is a $DCC1545M$ by
  ThorLabs}\BibitemShut {NoStop}%
\end{thebibliography}

\end{document}